# Hierarchical interfacial anchoring for durable perovskite solar cells

Jiaxin Wu[1,2], Hongzhuo Wu[1]*, Zuhong Zhang[1], Xiaoyu Guan[1], Jinbo Zhao[1], Yuting Guo[3], Zhihao Li[1], Zirui Li[1], Tengyu Ma[1], Ying Tang[1], Zhe Li[4], Weiwei Zuo[5], Jorge Pascual[6,7], Bo Hou[8], Feng Yang[9], Xiaoli Pan[10], Yu Wang[11], Luyao Wang[12]*, Lin-Long Deng[12], Guixiang Li[13], Luca Sapienza[3]*, Antonio Abate[2,14,15]*, Meng Li[1,3]*

**Affiliations:**

[1] Key Lab for Special Functional Materials of Ministry of Education, State Key Laboratory of Green Chemical Synthesis and Conversion, School of Nanoscience and Materials Engineering, Henan University, Kaifeng 475004, P. R. China.

[2] Department of Chemical, Materials and Industrial Production Engineering, University of Naples Federico II. Naples, pzz.le Vincenzo Tecchio 80, 80125 Naples, Italy.

[3] Department of Engineering, University of Cambridge, Cambridge, CB3 0FA, United Kingdom.

[4] School of Engineering and Materials Science (SEMS), Queen Mary University of London, London, E1 4NS, United Kingdom.

[5] Institute for Photovoltaics, University of Stuttgart, Pfaffenwaldring 47, 70569 Stuttgart, Germany.

[6] Instituto de Tecnologia Quimica (ITQ), Consejo Superior de Investigaciones Cientificas-Universitat Politecnica de Valencia, 46022, Valencia, Spain.

[7] Polymat, University of the Basque Country UPV/EHU, 20018 Donostia-San Sebastian, Spain.

[8] School of Physics and Astronomy, Cardiff University, Cardiff, Wales CF24 3AA, United Kingdom.

[9] College of Physics, Henan Key Laboratory of Advanced Semiconductor & Functional Device Integration, Henan Normal University, Xinxiang 453007, P.R. China.

[10] School of Intelligent Manufacturing (Pen-Tung Sah Institute of Micro-Nano Science and Technology), Xiamen University, Xiamen 361005, P. R. China

[11] New Energy Materials and Devices Laboratory, College of Materials and Chemistry, China Jiliang University, Hangzhou 310018, P. R. China.

[12] State Key Laboratory for Physical Chemistry of Solid Surfaces, Department of Chemistry, College of Chemistry and Chemical Engineering, Pen-Tung Sah Institute of Micro-Nano Science and Technology, Xiamen University, Xiamen 361005, P. R. China.

[13] School of Materials Science and Engineering, Southeast University, Nanjing 211189, P. R. China.

[14] Helmholtz-Zentrum Berlin für Materialien und Energie GmbH, Hahn-Meitner-Platz 1, 14109 Berlin, Germany.

[15] Department of Chemistry, Bielefeld University, Universitätsstraße 25, 33615 Bielefeld, Germany.

* Corresponding authors. Email: mengli@henu.edu.cn, ml2209@cam.ac.uk (M.L.); antonioabate83@gmail.com (A.A.); LS2052@cam.ac.uk (L.S.); wangluyaoyz@gmail.com (L.W.); wuhongzhuo@henu.edu.cn (H.W.)

**Abstract:**

Although inverted perovskite solar cells now deliver certified efficiencies approaching their single-junction limits, their practical deployment remains constrained by the thermo-mechanical fragility of buried interfaces under realistic temperature fluctuations. Here, we identify thermo-mechanically driven, stress-accumulation-dominated adhesion failure at self-assembled monolayer-based buried interfaces as a critical failure mode in inverted perovskite solar cells. To address this issue, we propose an interfacial robustness design strategy based on a homometallic halide pretreatment, which constructs a chemically compatible and highly anchorable interfacial environment on transparent conductive oxides to guide the ordered assembly of SAMs. This interfacial treatment increases the density of surface hydroxyl species and enables hierarchical multipoint interfacial anchoring dominated by metal-oxygen coordination, with $Cl^-$ induced hydrogen-bonding interactions acting cooperatively, thereby enhancing anchoring robustness and improving tolerance to thermo-mechanical stress accumulation. With this interfacial strategy, the devices achieve a power conversion efficiency of 27.3% (26.9% certified) and retain 94% of their initial efficiency after 400 temperature cycles between –40 and 85°C (ISOS-T-3). These results demonstrate that rational regulation of buried interfacial anchoring architectures can simultaneously enhance device efficiency and long-term operational reliability, offering a viable pathway toward high-efficiency and durable perovskite photovoltaics.

**One-Sentence Summary:**

Hierarchical interfacial anchoring enables certified 26.9% inverted perovskite solar cells with robust output under 1-sun illumination and thermal cycling from −40 to 85 °C.

## INTRODUCTION:

Perovskite solar cells (PSCs) have emerged as one of the most promising candidates for next-generation photovoltaic technologies owing to their high-power conversion efficiencies (PCE), low-temperature processing, and compatibility with tandem device architectures(*1-5*). In recent years, device efficiencies have exceeded 27%(*6-8*), further underscoring their potential for scalable applications. As this technology moves closer to practical deployment, achieving long-term operational reliability under realistic operating conditions has increasingly become a central challenge limiting its further development(*9*). Under real-world outdoor operation, temperature fluctuations introduce thermo-mechanical stresses that are fundamentally distinct from isothermal aging(*10*). During temperature cycling, mismatches in thermal expansion among functional layers generate cyclic strain incompatibilities(*9, 11-13*), inducing interfacial shear and sliding that lead to the progressive accumulation and amplification of interfacial stress, ultimately weakening adhesion and driving interfacial delamination, even in the absence of significant chemical degradation. This thermo-mechanically driven, stress-accumulation-dominated interfacial failure mode is particularly pronounced in inverted perovskite solar cells, where mechanically dissimilar layers are tightly coupled by ultrathin interfacial modifiers.

In inverted architectures, self-assembled monolayers (SAMs) have been widely used in high-efficiency perovskite solar cells owing to their excellent hole selectivity and molecular tunability(*14-19*). A typical SAM comprises a conjugated backbone and a phosphonic-acid anchoring group that self-assembles on transparent conductive oxides (TCOs) via P-O-M (M: In/Sn) coordination(*20*). Recent studies have converged on a consistent understanding of phosphonic acid-based SAM anchoring at oxide interfaces, showing that control of surface hydroxyl density and multidentate coordination enables dense, uniform, and chemically stable SAM layers(*15, 17, 21, 22*), while also revealing the sensitivity of anchoring configurations to long-term thermal stress and their implications for device reliability(*23*).

Against this backdrop, a physically relevant scenario under realistic operating conditions emerges: unavoidable temperature fluctuations subject even chemically stabilized SAM-based buried interfaces to repeated thermo-mechanical loading. Unlike isothermal degradation dominated by chemical aging, temperature cycling gives rise to interfacial stress accumulation through repeated shear and sliding**,** which can dominate failure. Under such conditions, even chemically robust SAM-oxide interfaces may undergo mechanically driven debonding as stresses accumulate. However, the evolution of this stress-accumulation-driven failure process, as well as the corresponding interfacial chemical design principles, remain largely unexplored**.**

On this basis, we identify thermo-mechanically driven debonding governed by interfacial stress accumulation at SAM-based buried interfaces as a critical failure mode in inverted perovskite solar cells and propose a homometallic halide pretreatment strategy to address this challenge. This strategy involves an in-situ hydrolysis process on indium tin oxide surfaces that increases the density of reactive hydroxyl species and establishes a chemically compatible interfacial template to promote ordered SAM assembly(*24-26*). The resulting SAMs adopt hierarchical multipoint anchoring configurations dominated by metal-oxygen coordination and supplemented by secondary interactions, thereby mitigating interfacial stress accumulation and enhancing tolerance to thermo-mechanical perturbations**.** Consequently, interfacial energy losses are suppressed, hole extraction is facilitated, and the resistance of buried interfaces to thermo-mechanical debonding is markedly improved. With this interfacial engineering strategy, we achieve a power conversion efficiency of 27.3% (26.9% certified), with 94% of the initial efficiency retained after 400 temperature cycles between −40 and 85 °C (ISOS-T-3). This work establishes the regulation of interfacial stress accumulation as a key design criterion for inverted perovskite photovoltaic devices, providing a scalable pathway toward simultaneously achieving high efficiency and long-term reliability.

## Thermo-mechanical failure of rigid, non-compliant interfacial anchoring at SAM-based buried interfaces.

As schematically illustrated in **Fig. 1A**, conventional SAM-based buried interfaces rely on rigid, non-compliant interfacial anchoring configurations, which offer limited tolerance to thermo-mechanical perturbations and are therefore susceptible to stress accumulation under temperature cycling. To elucidate the impact of temperature cycling on the perovskite/substrate interface, we performed accelerated aging following the International Summit on Organic Photovoltaic Stability (ISOS) protocol and conducted ISOS-T-3 thermal cycling (−40 to 85 °C)(*10*). After 100 temperature cycles, cross-sectional analysis revealed pronounced interfacial delamination at the buried contact (**Fig. 1B**). To further clarify the locus of interfacial failure, we analyzed both fracture surfaces of unaged and thermally cycled ITO/SAM/perovskite stacks by X-ray photoelectron spectroscopy (XPS) (**Fig. S1 and Supplementary Note 1**). After thermal cycling, no resolvable P 2p signal is detected on the ITO-side surface, whereas a clear P 2p signal is observed on the corresponding perovskite-side surface, indicating preferential transfer of the SAM with the perovskite during fracture. In contrast, unaged samples exhibit either cohesive fracture within the perovskite bulk or fracture at the buried interface, with the SAM remaining predominantly attached to ITO in the latter case. These contrasting fracture behaviors are consistent with preferential weakening of the SAM/ITO interface during thermal cycling.

Thermal cycling also alters the integrity of the SAM layer itself. SAM-only films subjected to identical aging conditions exhibit pronounced molecular aggregation in photoluminescence (PL) mapping, accompanied by a substantial reduction in surface coverage(*25*) (**Fig. 1C** and **Fig. S2**). In parallel, double-cantilever-beam (DCB) measurements show a marked deterioration of SAM/ITO adhesion(*12, 27, 28*), with the fracture energy decreasing from 1.59 to 0.50 J $m^{-2}$ after aging, a 69% reduction (**Fig. 1D**).

Taken together, these observations show that temperature cycling progressively degrades SAM integrity and buried-interface adhesion, with preferential weakening of the SAM/ITO contact emerging as a major contributor to the observed interfacial failure. Despite their chemical stability under isothermal conditions, phosphonic-acid SAMs anchored through rigid coordination exhibit limited tolerance to cyclic thermo-mechanical loading, leaving the buried interface vulnerable to mechanically driven debonding under realistic operating conditions.

## Interfacial regulation enables uniform SAM assembly and optimized energy alignment

Motivated by the mechanically driven interfacial failure identified above, we next sought to enhance the thermo-mechanical robustness of the buried interface by rationally engineering the anchoring architecture of SAMs. Accordingly, we develop a substrate modification strategy based on a homometallic halide pretreatment. This pretreatment induces in situ hydrolysis on the ITO surface (**Supplementary Note 2**), increasing the density of reactive hydroxyl species while introducing chemically compatible halide ligands. Concurrently, $In^{3+}$ species, chemically compatible with ITO because indium is a native lattice constituent, mitigate In-site defects and passivate lattice imperfections at the oxide surface. As a result, SAM molecules establish multipoint anchoring to the substrate through cooperative metal-ligand coordination and hydrogen-bonding interactions, forming a hierarchical anchoring architecture that accommodates interfacial strain under thermal cycling (**Fig. 2A**).

We first evaluated the optical impact of the $InCl_3$ layer. UV-vis transmission spectra show that the transmittance of the modified films is nearly identical to the control (**Fig. S3**), indicating excellent optical transparency. The treated substrate also exhibits reduced surface roughness and improved morphological uniformity, which favors continuous SAM coverage (**Fig. S4**)(*29*). XPS reveals a pronounced increase in surface hydroxyl content on ITO after $InCl_3$ treatment (**Fig. S5**) (24), which is primarily attributed to the formation of $In(OH)_3$ during in situ hydrolysis. This process simultaneously introduces abundant -OH and $Cl^-$ species at the surface. The -OH groups are covalently bound to ITO and thus exhibit greater stability than physisorbed species in polar solvents, whereas $Cl^-$ can be retained through coordination interactions with surface In sites. Together, these interactions effectively suppress SAM desorption during polar processing. The resulting high density of chemically stable anchoring sites provides a robust interfacial foundation for the formation of dense and well-anchored SAM layers.

To further clarify how the interfacial chemical environment is regulated, supplementary temperature-dependent XPS analyses show that the annealing temperature systematically modulates the surface hydroxyl density, oxygen-vacancy concentration, and halide retention on ITO (**Figs. S5, S6** and **Supplementary Note 3**). Intermediate annealing temperatures provide an optimal balance among hydroxyl availability, halide coordination stability, and oxide lattice integrity, yielding a surface chemistry most conducive to robust SAM anchoring.

On this basis, we next examine how such interfacial regulation translates into the uniformity, coverage continuity, and electronic alignment of the assembled SAMs. To evaluate the uniformity of the self-assembled monolayer, SAMs were deposited on both pristine and modified ITO substrates and subsequently characterized by XPS. The In-O-P/In-OH ratio increases after surface modification (**Fig. 2B**), consistent with enhanced phosphonate-related interfacial bonding enabled by the enriched anchoring sites(*21*). Furthermore, Kelvin probe force microscopy (KPFM) was employed to probe the surface potential distribution. The concomitant decrease in RMS roughness (**Fig. S7**) further corroborates the more homogeneous SAM formation(*14*).

In addition, ultraviolet photoelectron spectroscopy (UPS) reveals that the work function of the modified SAM aligns more closely with the valence-band maximum (VBM) of the perovskite, resulting in a deeper Fermi level and a reduced energetic offset(*30, 31*) (**Figs. 2D** and **S8**). This optimized energy-level alignment facilitates more efficient hole extraction while suppressing electron injection, thereby mitigating non-radiative recombination at the interface.

**Enhanced interfacial robustness under thermal cycling**

To elucidate the mechanistic origin of the enhanced interfacial robustness under thermal cycling, we combined theoretical analysis and experimental characterization to investigate the anchoring behavior of SAMs on the substrate (*18, 32*). XPS measurements reveal an overall positive shift of the In 3d core level, indicating reduced local electron density on In atoms. Together with the concomitant variations in the O 1s and P 2p spectra, these results are consistent with the formation of P-O-In interfacial interactions mediated by phosphonic groups, which contribute to an enhanced interfacial dipole (**Figs. S9** and **S10**). Collectively, the observed binding-energy shifts indicate strengthened interfacial electronic coupling and a more ordered, robust anchoring of SAMs on the modified substrate.

To evaluate the generality of the substrate modification strategy across different SAM molecules, MeO-2PACz and m-PhPACz were selected as representative systems, and their adsorption configurations on pristine and $InCl_3$-modified ITO surfaces were systematically constructed and compared. DFT calculations further elucidate the intrinsic mechanism by which cooperative primary and secondary anchoring enhance interfacial robustness. On pristine ITO, the absence of effective secondary anchoring pathways causes both SAMs to rely predominantly on rigid primary anchoring dominated by P-O-In interactions, resulting in mono- or bidentate adsorption configurations and hindering the stabilization of an ideal tridentate anchoring network (**Fig. 3A**, **S11**). In contrast, $InCl_3$ modification introduces $Cl^-$-mediated secondary anchoring, wherein surface $Cl^-$ species coordinate with undercoordinated In atoms and form hydrogen-bonding interactions with residual P-OH groups (**Fig. 3B**). Together with the primary P-O-In coordination, these interactions establish a hierarchical anchoring architecture (see **Supplementary Note 4** for definitions of the anchoring terminology used throughout this work). This secondary anchoring cooperates with the P-O-In primary anchoring to stabilize a tridentate anchoring configuration. Consequently, the adsorption energies of both SAMs are substantially enhanced after $InCl_3$ treatment, by 0.61 eV for m-PhPACz and 1.28 eV for MeO-2PACz (**Fig. 3C**, **S12**). The enhancement observed for both SAMs supports the applicability of this $InCl_3$ surface-engineering strategy, based on a hierarchical primary-secondary anchoring framework, to the representative phosphonic-acid SAMs studied here.

Importantly, the synergistic interplay between the rigid primary anchoring and the compliant secondary anchoring may provide the tridentate anchoring network with a hydrogen-bond-mediated pathway for reversible interfacial recovery. The primary anchoring, dominated by P-O-In coordination/covalent interactions, offers high bond energy and thermal stability but is less likely to undergo spontaneous recovery once disrupted: when an In-O(P) bond ruptures under locally concentrated thermomechanical stress, the resulting undercoordinated sites are unlikely to spontaneously re-form strong interfacial bonds. In contrast, the secondary anchoring involving P-OH···Cl hydrogen bonds may undergo dissociation and re-formation under thermal perturbation, potentially re-establishing local interfacial contacts and providing a pathway for dynamic interfacial recovery (see **Fig. S14**, **Supplementary Note 5** for details). Accordingly, this hierarchical anchoring architecture may facilitate interfacial stress dissipation and local recovery during cyclic loading, thereby contributing to enhanced anchoring stability and interfacial toughness.

To further assess anchoring stability, we quantified SAM coverage by XPS before and after thermal cycling (**Fig. S15**). After 100 cycles, the coverage factor decreased only slightly from $30.6\times10^{-3}$ to $29.5\times10^{-3}$; EDS corroborates this observation (**Fig. S16**). C-AFM imaging further reveals a clear contrast in interfacial evolution upon thermal cycling (**Fig. S17**): the control substrate shows a marked increase in non-conductive regions after cycling, corroborating the deterioration of SAM/ITO interfacial integrity indicated by XPS, whereas the target substrate remains largely conductive, with only a modest increase in the non-conductive fraction. PL mapping (**Fig. 3D**) shows no discernible change in the SAM film after thermal cycling, consistent with robust multipoint interfacial anchoring. Accordingly, the interfacial mechanical integrity is substantially enhanced by the optimized interfacial design and is well retained after cycling (**Fig. 3E**). Cross-sectional imaging in **Fig. S18** further confirms good perovskite/substrate contact after cycling, validating the robustness of the substrate treatment(*33*).

Finally, SEM and XRD indicate that substrate optimization reduces voids and enhances crystallinity in the perovskite layer, despite similar grain sizes, consistent with more uniform SAM coverage (**Figs. S19** and **S20**)(*34*). Grazing-incidence X-ray diffraction (GIXRD) shows that the control film retains substantial in-plane tensile residual stress after thermal cycling, whereas this stress is markedly relieved in the target film (**Figs. 3F-H**), consistent with the role of native $In^{3+}$ in defect passivation and the alleviation of local residual stress(*35*).

## Improved charge dynamics and photovoltaic performance

To probe defect states in the perovskite, we performed space-charge-limited current (SCLC) measurements (**Fig. S21**), which show that the substrate modification significantly reduces the defect density(*34*). Light-intensity-dependent $V_{OC}$ measurements (**Fig. S22**) were used to analyze transport and recombination kinetics; the optimized devices exhibit suppressed nonradiative recombination(*36*). Dark-current measurements reveal lower leakage in target devices (**Fig. S23**), likely due to improved interfacial contact and enhanced hole extraction. Consistently, electrochemical impedance spectroscopy (EIS) indicates a lower defect density and improved charge-transport capability (**Fig. S24**), consistent with the kinetic improvements above(*37*).

In the interfacial analyses above, the widely used MeO-2PACz was employed as a representative SAM. Our calculations further reveal that the modification strengthens the adsorption of both MeO-2PACz and m-PhPACz on the ITO substrate, demonstrating the broad applicability of this interfacial strategy across different SAM chemistries. In line with these computational findings, devices based on both SAMs exhibit performance improvements after $InCl_3$ treatment, with increases in $V_{OC}$ and FF (**Figs S25** and **S26**). Owing to its more favorable energetic alignment, larger steric hindrance, and stronger interactions with the perovskite absorber, which collectively result in superior device performance after $InCl_3$ modification, m-PhPACz was selected for the champion device and stability assessments.

Devices are configured in a p-i-n stack of ITO/SAM/perovskite/interlayer/$C_{60}$/$SnO_X$/Ag (**Fig. S27**). Under AM 1.5G illumination, the $InCl_3$-treated device shows superior J-V characteristics compared with the control (**Fig. 4A**), reaching a PCE of 27.3%. The target device also shows markedly reduced *J-V* hysteresis relative to the control, with a hysteresis index decreasing from 4.3% to 2.2%. Notably, $V_{OC}$ increases from 1.18 V to 1.21 V, and FF from 82.9% to 85.4%, reflecting reduced energy losses. The EQE-integrated $J_{SC}$ is 26.1 mA/cm$^2$, consistent with *J-V*; EQE derivative analysis yields a bandgap of 1.53 eV (**Fig. S28**). The champion device with an active area of 7.81 mm² is certified at 26.9% by a national PV metrology center (**Figs. 4B, S29, S30**) and reaches a steady-state PCE of 26.5%, well above the control (24.5%, **Fig. 4c**). To evaluate the scalability of the $InCl_3$ modification strategy, devices with different active areas were further examined. The 1 cm$^2$ device achieves a PCE of 25.7%, and even at a significantly enlarged area of 70 cm² the device maintains an efficiency of 22.7% (**Fig. S31, S32**), demonstrating strong potential for module-scale applications.

To directly assess interfacial energetic losses at the buried contact, we conducted QFLS mapping on ITO/SAM/perovskite stacks. The average QFLS increases from 1.20 ± 0.03 to 1.23 ± 0.02 eV after modification (**Figs. 4D, 4F**), evidencing reduced energy loss and improved spatial uniformity(*38-41*). We attribute these gains to the uniform, dense SAM coverage and more robust interfacial coupling enabled by the substrate treatment.

## Operational stability of the devices

We systematically evaluated device operational stability following the ISOS protocols. We first performed rapid thermal cycling (ISOS-T-3, −40 to 85 °C, 18 min per cycle). As shown in **Fig. 4G**, the control device retained only 61% of its initial PCE after 400 cycles, whereas the target device maintained 94%, demonstrating excellent tolerance to large temperature excursions. Under continuous illumination with maximum power-point tracking (MPPT) (**Fig. 4H**), the target device retained 95% of its initial PCE after 2000 h. Furthermore, we evaluated the thermal stability of the devices following the ISOS-D-2 protocol(*10*). The target device retained 95% of its initial efficiency after 1000 h of heating, whereas the control exhibited pronounced degradation (**Fig. S33**). The devices also showed robust storage stability (**Fig. S34**). Taken together, these results indicate that the interfacial pretreatment effectively suppresses SAM desorption and thermo-mechanical damage at the buried contact during thermal cycling, thereby markedly enhancing operational stability.

## DISCUSSION

In summary, thermal cycling induces interfacial stress accumulation through repeated thermo-mechanical shear and sliding, which ultimately drives adhesive failure at buried interfaces. This failure process is effectively mitigated by the homometallic halide pretreatment introduced here. This pretreatment, combined with in situ hydrolysis, increases the density of available anchoring sites (-OH/$Cl^-$) and establishes a stable, continuous interfacial template for SAM assembly, enabling uniform and robust SAM deposition while reducing energetic losses and facilitating efficient charge separation and extraction. On this engineered interface, phosphonic SAMs form a hierarchical tridentate anchoring network through cooperative rigid primary anchoring (P-O-In interactions) and compliant secondary anchoring ($Cl^-$-mediated hydrogen bonding). The resulting bonding redundancy and the potentially reorganizable nature of the secondary hydrogen-bonding interactions may assist the re-establishment of local interfacial contacts under thermal perturbation, thereby contributing to enhanced interfacial mechanics and long-term device operation. With this interfacial chemistry, champion devices achieve a power conversion efficiency of 27.3%, with a certified efficiency of 26.9%. This work establishes hierarchical interfacial anchoring as a general design principle for mitigating thermo-mechanically driven interfacial adhesion failure, enabled here by homometallic metal-halide surface engineering, and provides a scalable pathway toward durable and high-efficiency perovskite photovoltaics.

## MATERIALS AND METHODS

### Materials

Lead iodide ($PbI_2$) and fullerene ($C_{60}$) were purchased from Libra Technology Corporation. Formamidinium iodide (FAI) and methylammonium iodide (MAI) were purchased from Greatcell Solar. [2-(3,6-Dimethoxy-9H-carbazol-9-yl) ethyl] phosphonic Acid (MeO-2PACz), piperazine dihydriodide ($PDI_2$) and cesium iodide (CsI) were purchased from TCI. Organic solvents were purchased from Alfa Aesar, including dimethyl sulfoxide (DMSO, 99.8%) and N, N-dimethylformamide (DMF, 99.8%). Chlorobenzene (CB, 99.5%), isopropanol (IPA, 99.7%), and absolute ethyl alcohol. Indium chloride ($InCl_3$) was purchased from Macklin. The m-PhPACz molecule was prepared according to the synthetic protocol described in our previous publication.

### Device Fabrication

Solution preparation: An $InCl_3$ solution (1 mg $mL^{-1}$) was prepared in ultrapure water and stirred at room temperature for 1 h. Two separate SAM solutions were prepared by dissolving MeO-2PACz and m-PhPACz individually in ethanol at 0.5 mg/mL. Each solution was stirred at 25 °C for 6 h and used without filtration. The perovskite precursor solution was prepared by dissolving 1.7 mmol of $Cs_{0.05}MA_{0.05}FA_{0.9}PbI_3$ in 1 mL of a mixed DMF/DMSO solvent (4:1 v/v). The solution was stirred at 60 °C for 6 h and filtered through a 0.22 μm hydrophobic PTFE syringe filter before use. $PDI_2$ (0.4 mg) was dissolved in 1 mL isopropanol and stirred at 25 °C for 12 h before use, without filtration.

Device preparation: Indium tin oxide (ITO) glasses were cleaned using detergent, followed by sequential sonication in deionized water, acetone, and IPA (10 min each). The substrates were dried with $N_2$ and treated with UV-ozone for 30 min before being transferred into a glovebox. For the target device, an $InCl_3$ layer was deposited by spin-coating at 3000 rpm for 30 s and annealing at 130 °C for 10 min. For the control device, this $InCl_3$ step was omitted, while all subsequent fabrication steps were kept identical. The SAM layer was formed by spin-coating 100 μL of the corresponding SAM solution at 3000 rpm for 30 s and annealing at 100 °C for 10 min. The perovskite precursor solution (100 μL) was deposited by a two-step spin-coating process at 1000 rpm for 10 s and 5000 rpm for 30 s, during which 200 μL of chlorobenzene was dripped at 25 s. The film was then annealed at 100 °C for 45 min. The $PDI_2$ solution (100 μL) was spin-coated at 5000 rpm for 30 s and annealed at 100 °C for 10 min. Finally, a 35 nm $C_{60}$ layer, a 20 nm $SnO_X$ layer, and a 100 nm Ag electrode were sequentially deposited through a shadow mask. $SnO_X$ was deposited by thermal atomic layer deposition (ALD), while $C_{60}$ and Ag were deposited by thermal evaporation under vacuum.

Fabrication of modules: Perovskite solar cell modules with an aperture area of 70 $cm^2$ were fabricated on 10.5 × 10.5 $cm^2$ ITO-coated glass substrates. Sub-cells were monolithically connected in series by laser scribing (P1, P2, and P3). The P1 scribe (30 μm) was performed on ITO using 16 W power, 100 ns pulse width, and 120 kHz repetition rate. P2 and P3 scribes were conducted with 5 W power, 100 ns pulse width, and 80 kHz repetition rate. The overall geometric fill factor (GFF) was approximately 96%. After P1, the ITO substrates were cleaned by sequential ultrasonic washing in detergent, acetone, and IPA (each for 10 min), dried under $N_2$, and treated with UV-ozone for 30 min. An $InCl_3$ layer was deposited on the glass. For the hole-selective interface, a m-PhPACz solution was deposited by immersing the substrates in a 0.5 mg/mL ethanolic solution of SAM for 10 minutes, followed by annealing at 100°C for 10 minutes. The substrates were then rinsed with ethanol and dried with nitrogen. The perovskite precursor solution contained perovskite at a total concentration of 1.4 M dissolved in DMF: DMSO (4:1, v/v). The solution was spin-coated onto the SAM-treated ITO, a coating speed of 4000 rpm for 35 s, at 25 s, adding 1000 μL of CB. The films were then annealed at 100 °C for 45 min in $N_2$. The interface passivation was the same of the small-scale devices. Next, 40 nm of $C_{60}$ was thermally evaporated onto the perovskite layer at a deposition rate of 0.3-0.5 Å/s. The P2 scribe (80 μm) was then performed as above, and 20 nm of $SnO_x$ was deposited by atomic layer deposition. Finally, the top electrode consisting of 100 nm Ag was thermally evaporated, followed by the P3 scribe (60 μm).

## Characterization

Device performance: Current density-voltage (*J-V*) characteristics were measured under AM 1.5G illumination using a solar simulator (EnliTech SS-X50, A+ spectrum) coupled with a Keithley 2400 source meter. The light intensity was calibrated against a NIST-traceable silicon reference cell. *J-V* scans were acquired at 10 mV/s for small-area devices and 50 mV/s for modules, using a metal aperture in a glovebox. The integrated $J_{SC}$ from the external quantum efficiency (EQE) was obtained using a solar-cell spectral response system (EnliTech QE-R). Device stability was evaluated under LED illumination at 100 mW $cm^{-2}$, with the intensity calibrated using a standard silicon cell; all measurements were conducted in an inert glovebox atmosphere. For thermal-cycling tests, the device was placed in a chamber with controlled vacuum and temperature and cycled between −40 °C and 85 °C, with liquid nitrogen as the coolant. For thermal stability tests, devices were aged on a hot plate at 85 °C under nitrogen.

Mechanical Testing: The “sandwich” double-cantilever beam (DCB) specimens were prepared with the following structure: glass/ITO/SAM/Perovskite/PMMA/epoxy/ITO/glass(*12*). The SAM and perovskite layers were deposited following the procedures described above. To probe the role of $InCl_3$, additional samples incorporating $InCl_3$ were prepared. An ~800-nm PMMA protection layer was formed on the MHP film by spin-coating a 10 wt% PMMA solution in chlorobenzene at 4000 rpm for 40 s and drying at room temperature for 1 h. A ~2-μm epoxy layer was then applied onto the PMMA to bond a second, cleaned ITO-coated glass substrate. The PMMA interlayer was used to prevent adverse reactions between the epoxy and the perovskite film. Before epoxy application, a ~5-mm region along the short edge was masked to create a notch. The DCB specimens were cured in a drybox for 24 h, and excess epoxy was carefully removed from the edges prior to mechanical testing.

Perovskite film properties: Surface morphology was characterized by top-view SEM, (Zeiss Supra 55, SE2 mode, 5 kV). XPS was carried out using a Thermo Fisher ESCALAB 250Xi, and the resulting spectra were analyzed with Thermo Avantage. UV-Vis absorption spectra were collected using a Cary 5000 spectrophotometer, and FTIR spectra were obtained on a Bruker VERTEX 70 system. Surface potential mapping was conducted using AFM, (SPA400, Zeiss). UPS measurements were carried out on a Kratos Axis Ultra system with a monochromatic Al Kα source and He-I excitation (21.22 eV) under ultra-high vacuum. GIXRD and conventional XRD of perovskite films were measured using a Rigaku SmartLab 3 kW diffractometer and a Bruker Discovery D8 diffractometer (Cu Kα radiation), respectively.

## Theoretical calculation

Spin-polarized density functional theory calculations were performed using the Perdew-Burke-Ernzerhof (PBE)(*42*) functional and the projector augmented wave (PAW)(*43*) potential as implemented in the Vienna Ab Initio Simulation Package (VASP)(*44, 45*). An energy cutoff of 400 eV and a convergence criterion of $10^{-7}$ eV for self-consistent calculations was adopted. All structures were relaxed until the total force on each relaxed atom was less than 0.04 eV/ Å. The van der Waals interaction was corrected based on the DFT-D3 scheme(*46*). The thickness of the vacuum layer was larger2*2 than 15 Å. A Γ-centered 2 × 2 × 1 k-point mesh was adopted. VASPKIT code(*47*) and VESTA software(*48*) were used for calculation pre-processing and post-processing.

Classical molecular dynamics (MD) simulations were performed with GROMACS for SAMs on ITO substrate(*49, 50*). GAFF force field was used for SAMs and UFF force field was used for ITO. Periodic boundary conditions were applied to the xy-direction, and the ITO substrate was fixed. The adsorption model included a 76 Å × 76 Å × 14 Å modified $In_2O_3$ supercell as ITO. The system contained 20 MeO-2PACz. The REPEAT charges were used as the atomic charges for ITO, computed by CP2K. The restrained electrostatic potential (RESP) charges were used as the atomic charges for SAM, computed by ORCA and Multiwfn(*51, 52*). The forcefield parameters and topology files were created by Sobtop(*53*). The system was simulated for 10 ns (i.e., 10000 ps) under canonical ensemble (NVT) ensemble with a time step of 1 fs under the Velocity-rescale thermostat. The hydrogen bond analyses were finished by VMD program.

**Equations for calculation**

Analysis for Light intensity-dependent $V_{OC}$:

$$V_{\mathrm{OC}} = \frac{nK_B T lnI}{q} + C \qquad \text{(Equation 1)}$$

Where the n is Ideality factor, $K_B$ is Boltzmann's constant, T is Kelvin temperature, q is electron charge, which is usually between 1 and 2, the smaller n is theoretically, the indirect defect recombination at the interface (SRH recombination, Shockley-Read-Hall recombination) is smaller.

Analysis for the FF losses:

$$FF_{max} = \frac{v_{oc} - ln\,(v_{oc}+0.72)}{v_{oc}+1} \qquad \text{(Equation 2)}$$

$$v_{oc} = \frac{qV_{oc}}{nK_B T} \qquad \text{(Equation 3)}$$

Where $K_B$ is Boltzmann's constant, T is Kelvin temperature and q is electron charge, n is Ideality factor.

**Supplementary Materials**

Supplementary Text

Figs. S1 to S34

References

**References and Notes**

1. X. Zhang *et al.*, Advances in inverted perovskite solar cells. *Nat. Photon.* **18**, 1243–1253 (2024).
2. P. Chen *et al.*, The promise and challenges of inverted perovskite solar cells. *Chem. Rev.* **124**, 10623–10700 (2024).
3. S. Liu *et al.*, Buried interface molecular hybrid for inverted perovskite solar cells. *Nature* **632**, 536–542 (2024).
4. Y. Yang *et al.*, Amidination of ligands for chemical and field-effect passivation stabilizes perovskite solar cells. *Science* **386**, 898–902 (2024).
5. A. Kojima, K. Teshima, Y. Shirai, T. Miyasaka, Organometal halide perovskites as visible-light sensitizers for photovoltaic cells. *J. Am. Chem. Soc.* **131**, 6050–6051 (2009).
6. . (NREL. Best research-cell efficiency chart., 2025), vol. 2025.
7. G. Li *et al.*, Stabilizing high-efficiency perovskite solar cells via strategic interfacial contact engineering. *Nat. Photon.* **19**, 1123–1131 (2025).
8. Z. Xiong *et al.*, Homogenized chlorine distribution for> 27% power conversion efficiency in perovskite solar cells. *Science* **390**, 638–642 (2025).
9. G. Li *et al.*, Highly efficient pin perovskite solar cells that endure temperature variations. *Science* **379**, 399–403 (2023).
10. M. V. Khenkin *et al.*, Consensus statement for stability assessment and reporting for perovskite photovoltaics based on ISOS procedures. *Nat. Energy* **5**, 35–49 (2020).
11. Q. Jiang *et al.*, Surface reaction for efficient and stable inverted perovskite solar cells. *Nature* **611**, 278–283 (2022).
12. Z. Dai *et al.*, Interfacial toughening with self-assembled monolayers enhances perovskite solar cell reliability. *Science* **372**, 618–622 (2021).
13. B. Dong *et al.*, Self-assembled bilayer for perovskite solar cells with improved tolerance against thermal stresses. *Nat. Energy* **10**, 342–353 (2025).
14. H. Wu *et al.*, Tailored lattice-matched carbazole self-assembled molecule for efficient and stable perovskite solar cells. *J. Am. Chem. Soc.* **147**, 8004–8011 (2025).
15. H. Tang *et al.*, Reinforcing self-assembly of hole transport molecules for stable inverted perovskite solar cells. *Science* **383**, 1236–1240 (2024).
16. W. Wu *et al.*, Stable and uniform self-assembled organic diradical molecules for perovskite photovoltaics. *Science* **389**, 195–199 (2025).
17. S. Fu *et al.*, Homogenizing SAM deposition via seeding–OH groups for scalable fabrication of perovskite solar cells. *Energy Environ. Sci.* **18**, 3305–3312 (2025).
18. Z. Li *et al.*, Stabilized hole-selective layer for high-performance inverted pin perovskite solar cells. *Science* **382**, 284–289 (2023).
19. W. Jiang *et al.*, Toughened self-assembled monolayers for durable perovskite solar cells. *Nature* **646**, 95–101 (2025).
20. E. Aydin *et al.*, Enhanced optoelectronic coupling for perovskite/silicon tandem solar cells. *Nature* **623**, 732–738 (2023).

21. C. Luo *et al.*, Engineering bonding sites enables uniform and robust self-assembled monolayer for stable perovskite solar cells. *Nat. Mater.* **24**, 1265–1272 (2025).
22. H. Zhang *et al.*, A Self‐Assembling Composite Structural Design for the Conversion of Hydroxyl‐Anchored Bonds Obtains High Efficient and Stable Perovskite Solar Cells. *Adv. Mater.* **37**, 2420155 (2025).
23. C. Fei *et al.*, Limiting phosphonic acid interlayer–perovskite reactivity to stabilize perovskite solar modules. *Science* **391**, eadz7969 (2026).
24. J. Xiang, Z. X. Liu, H. Chen, C. Z. Li, Robust and sustainable indium anode leading to efficient and stable organic solar cells. *Adv. Mater.* **35**, 2303729 (2023).
25. X. Zhang *et al.*, Reinforcing Coverage of Self‐assembled Monomolecular Layers for Inverted Perovskite Solar Cells with Efficiency of 25.70%. *Angew. Chem. Int. Ed.* **64**, e202423827 (2025).
26. G. Liu *et al.*, Halogen Radical‐Activated Perovskite‐Substrate Buried Heterointerface for Achieving Hole Transport Layer‐Free Tin‐Based Solar Cells with Efficiencies Surpassing 14%. *Angew. Chem. Int. Ed.* **137**, e202419183 (2025).
27. Z. Li *et al.*, Hyperbranched polymer functionalized flexible perovskite solar cells with mechanical robustness and reduced lead leakage. *Nat. Commun.* **14**, 6451 (2023).
28. A. S. Bati *et al.*, A chemically bonded monolayer interface enables enhanced thermal stability and efficiency in Pb-Sn perovskite solar cells. *Joule* **9**, 102047 (2025).
29. R. Guo *et al.*, Refining the substrate surface morphology for achieving efficient inverted perovskite solar cells. *Adv. Energy Mater.* **13**, 2302280 (2023).
30. J. Wu *et al.*, Bisphosphonate‐anchored self‐assembled molecules with larger dipole moments for efficient inverted perovskite solar cells with excellent stability. *Adv. Mater.* **36**, 2401537 (2024).
31. G. Qu *et al.*, Conjugated linker-boosted self-assembled monolayer molecule for inverted perovskite solar cells. *Joule* **8**, 2123–2134 (2024).
32. C. Li *et al.*, Fully Aromatic Self‐Assembled Hole‐Selective Layer toward Efficient Inverted Wide‐Bandgap Perovskite Solar Cells with Ultraviolet Resistance. *Angew. Chem. Int. Ed.* **63**, e202315281 (2024).
33. J. Wen *et al.*, Buried heterointerface reinforcement with passivation-integrated nanostructures for efficient and stable perovskite solar modules. *Joule* **9**, 102212 (2025).
34. J. Wu *et al.*, Inhibiting interfacial nonradiative recombination in inverted perovskite solar cells with a multifunctional molecule. *Adv. Mater.* **36**, 2407433 (2024).
35. Y. Tang *et al.*, Heat‐Triggered Dynamic Self‐Healing Framework for Variable‐Temperature Stable Perovskite Solar Cells. *Adv. Mater.*, 2420378 (2025).
36. Z. Zhang *et al.*, Improved Air Stability of Tin Halide Perovskite Solar Cells by an N‐Type Active Moisture Barrier. *Adv. Funct. Mater.* **34**, 2306458 (2024).
37. Z. Zhang *et al.*, Anchoring Charge Selective Self‐Assembled Monolayers for Tin–Lead Perovskite Solar Cells. *Adv. Mater.* **36**, 2312264 (2024).
38. S. Zhumagali *et al.*, Efficient Narrow Bandgap Pb‐Sn Perovskite Solar Cells Through Self‐Assembled Hole Transport Layer with Ionic Head. *Adv. Energy Mater.* **15**, 2404617 (2025).

39. F. Xu *et al.*, Stabilized perovskite phases enabling efficient perovskite/perovskite/silicon triple-junction solar cells. *Nat. Mater.* **24**, 1365–1374 (2025).
40. H. Wang *et al.*, Yttrium oxide engineered substrate enables improved durability for perovskite solar cells. *Nat. Commun.* **16**, 9508 (2025).
41. O. Er-Raji *et al.*, Electron accumulation across the perovskite layer enhances tandem solar cells with textured silicon. *Science* **390**, eadx1745 (2025).
42. J. P. Perdew, K. Burke, M. Ernzerhof, Generalized gradient approximation made simple. *Phys. Rev. Lett.* **77**, 3865 (1996).
43. P. E. Blöchl, Projector augmented-wave method. *Phys. Rev. B* **50**, 17953 (1994).
44. G. Kresse, J. Furthmüller, Efficient iterative schemes for ab initio total-energy calculations using a plane-wave basis set. *Phys. Rev. B* **54**, 11169 (1996).
45. G. Kresse, J. Furthmüller, Efficiency of ab-initio total energy calculations for metals and semiconductors using a plane-wave basis set. *Comput. Mater. Sci.* **6**, 15–50 (1996).
46. S. Grimme, J. Antony, S. Ehrlich, H. Krieg, A consistent and accurate ab initio parametrization of density functional dispersion correction (DFT-D) for the 94 elements H-Pu. *J. Chem. Phys.* **132**, 154104 (2010).
47. V. Wang, N. Xu, J.-C. Liu, G. Tang, W.-T. Geng, VASPKIT: A user-friendly interface facilitating high-throughput computing and analysis using VASP code. *Comput. Phys. Commun.* **267**, 108033 (2021).
48. K. Momma, F. Izumi, VESTA 3 for three-dimensional visualization of crystal, volumetric and morphology data. *J. Appl. Crystallogr.* **44**, 1272–1276 (2011).
49. M. J. Abraham *et al.*, GROMACS: High performance molecular simulations through multi-level parallelism from laptops to supercomputers. *SoftwareX* **1**, 19–25 (2015).
50. B. Hess, C. Kutzner, D. Van Der Spoel, E. Lindahl, GROMACS 4: algorithms for highly efficient, load-balanced, and scalable molecular simulation. *J. Chem. Theory Comput.* **4**, 435–447 (2008).
51. T. Lu, F. Chen, Multiwfn: A multifunctional wavefunction analyzer. *J. Comput. Chem.* **33**, 580–592 (2012).
52. T. Lu, A comprehensive electron wavefunction analysis toolbox for chemists, Multiwfn. *J. Chem. Phys.* **161**, 082503 (2024).
53. W. Humphrey, A. Dalke, K. Schulten, VMD: visual molecular dynamics. *J. Mol. Graph.* **14**, 33–38 (1996).

## Acknowledgements:

**Funding:** This work was supported by the University of Cambridge, Henan University, Helmholtz-Zentrum Berlin für Materialien und Energie (HZB), and Xiamen University (XMU). The authors acknowledge financial support from the National Natural Science Foundation of China (General Program, Grant No. 52472199), the Key Project for International Science and Technology Cooperation of Henan Province (Grant No. 261111520200 (K26193Y)), the Program for the Science and Technology Innovation Team in Universities of Henan Province (Grant No. 26IRTSTHN001 (K26005Y)), the Key Research and Development and Promotion Projects of Henan Province (Grant No. 262102230144), and the Joint Fund of the Provincial Science and Technology Research and Development Plan of Henan Province (Grant No. 232301420004).

**Author contributions:** J.W., H.W., L.W., and M.L. conceived the idea. J.W., Z.Z., J.Z., X.G., Zirui L. and T.M. fabricated the devices and measured the data, H.W., L.W., L.S., A.A., and M.L. performed the data analysis and discussed the results. Zhihao L. and J.W. prepared the samples for DCB measurements and were responsible for the corresponding data analysis. X.P., and Y.W. performed the C-AFM measurement. In addition, J.W., L.W. and Z.Z. contributed to the stability evaluation of the devices by collecting and analyzing long-term performance data. J.W. and H.W. wrote the initial manuscript. Subsequent revisions were made by J.W., H.W., Y.G., Y.T., Z.L., W.Z., J.P., B.H., F.Y., L.W., L.-L.D., G.L., L.S., A.A., and M.L. H W., L.W., L.S., A.A., and M.L. provided expertise and supervised the work. All authors revised the final version.

**Competing interests:** The Authors declare that they have no competing interests.

**Data and materials availability:** All data and code needed to evaluate and reproduce the results in the paper are present in the paper and/or the Supplementary Materials. This study did not generate any new materials.

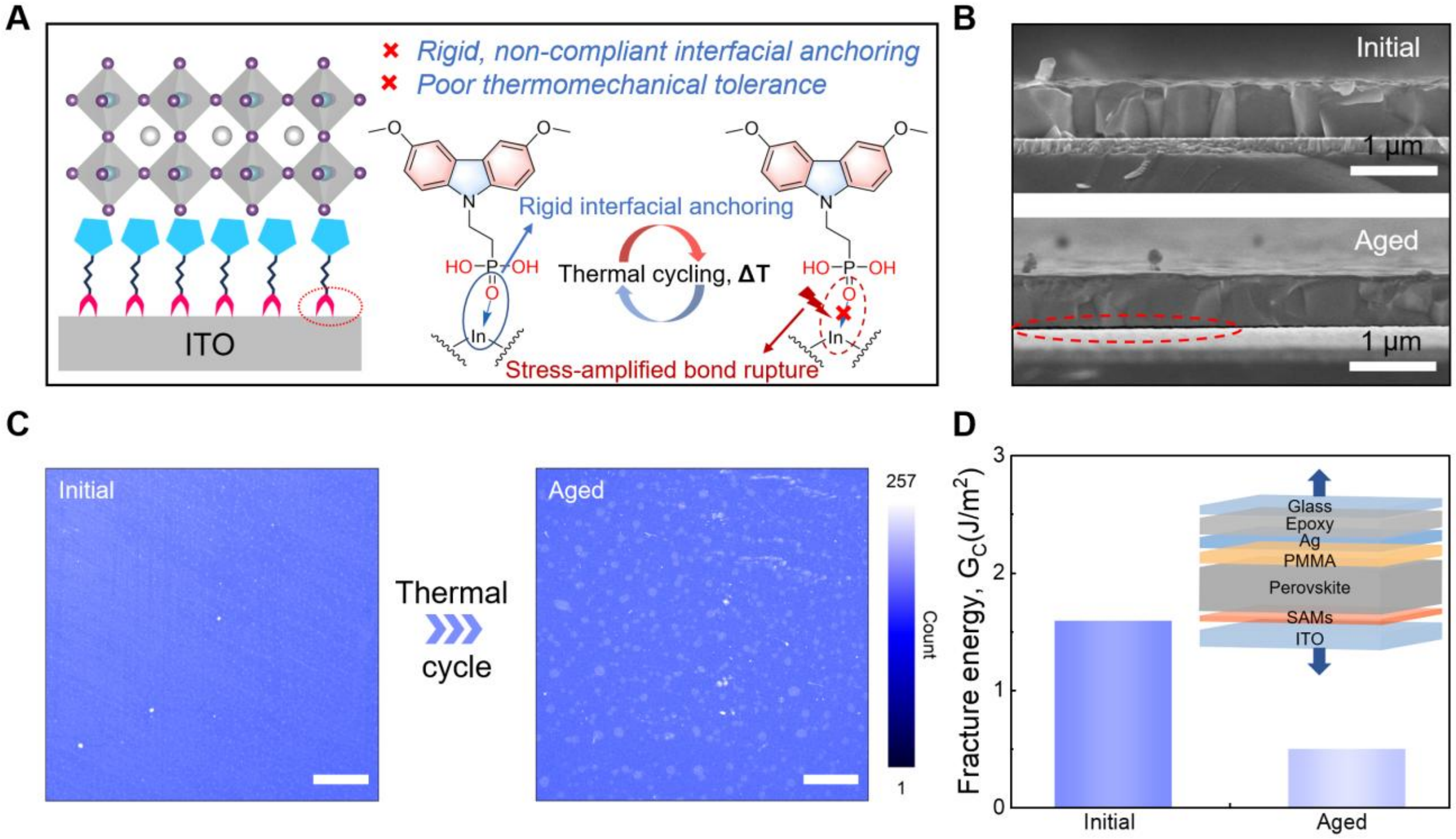

**Fig. 1. Thermo-mechanical failure of rigid, non-compliant interfacial anchoring at SAM-based buried interfaces.** (A) Schematic illustration of the thermo-mechanical instability of conventional SAM-based buried interfaces on ITO. Rigid, non-compliant interfacial anchoring dominated by single-site metal-oxygen bonding provides limited tolerance to thermal mismatch. Under thermal cycling (ΔT), thermomechanical shear and interfacial sliding lead to stress concentration at the rigid anchoring sites, resulting in stress-amplified bond rupture and progressive interfacial degradation. (B) Cross-sectional SEM images of the perovskite films before and after thermal cycling. (C) PL mapping images of the SAM films before and after thermal cycling. (D) Interfacial fracture energy (GC) extracted from double-cantilever-beam (DCB) tests for the perovskite/SAM/ITO stack in the initial and thermally aged states.

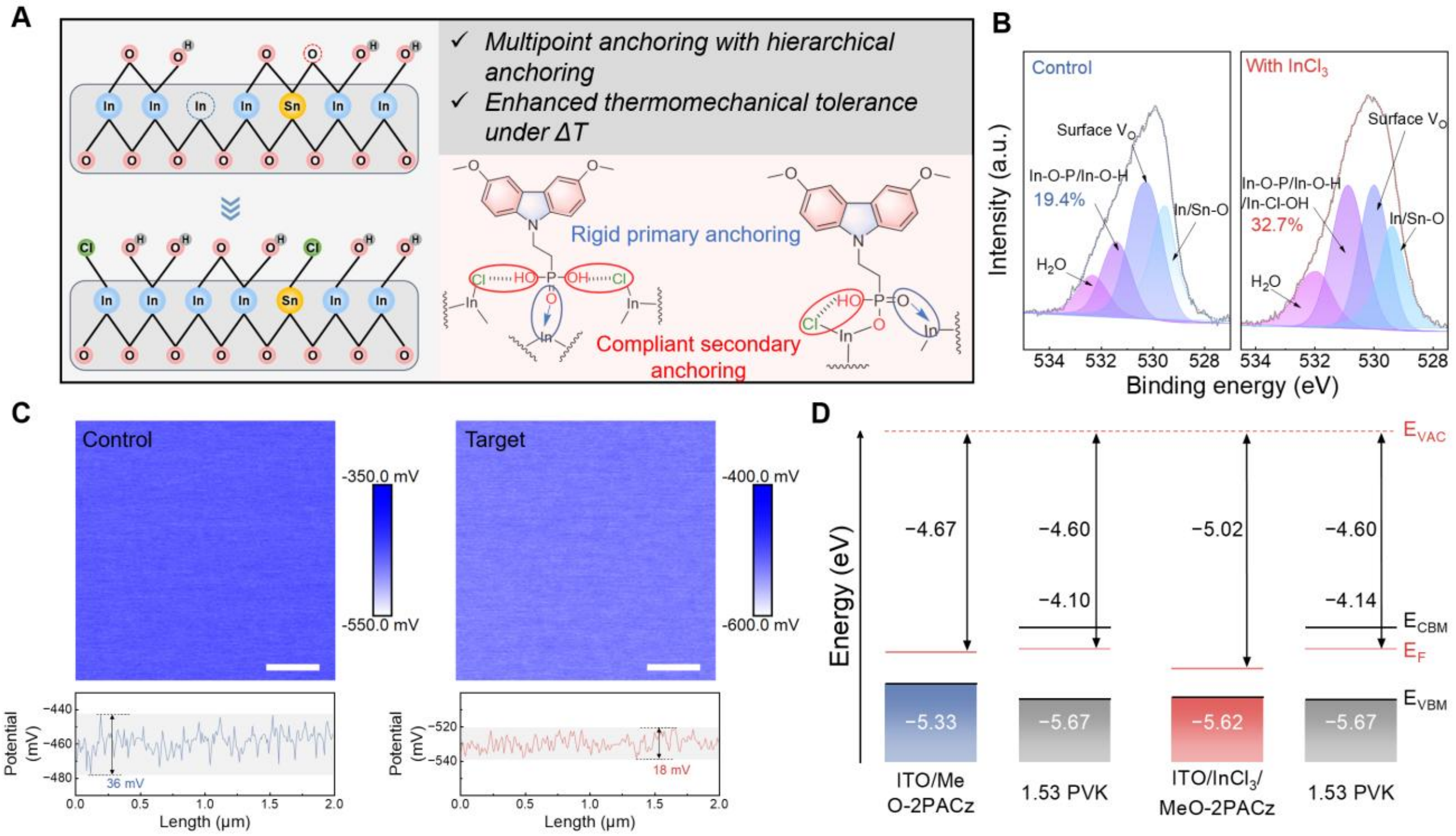

**Fig. 2. Providing ample anchoring sites for SAMs.** (**A**) Schematic illustration of the substrate modification process. (**B**) O 1s XPS spectra and peak deconvolution of SAMs assembled on the control and target substrates. The gray line shows the experimental XPS spectrum, while the dashed line denotes the overall fitted envelope; individual fitted components are shown separately. (**C**) KPFM surface potential maps and corresponding line profiles of the SAM-coated control and target substrates. (**D**) Schematic representation of the band-edge positions of the studied systems based on UPS measurements, referenced to the vacuum level. $E_F$ and $E_{VAC}$ represent Fermi and vacuum levels, respectively. $E_{VBM}$ and $E_{CBM}$ represent the energy of valence band maximum and conduction band minimum, respectively.

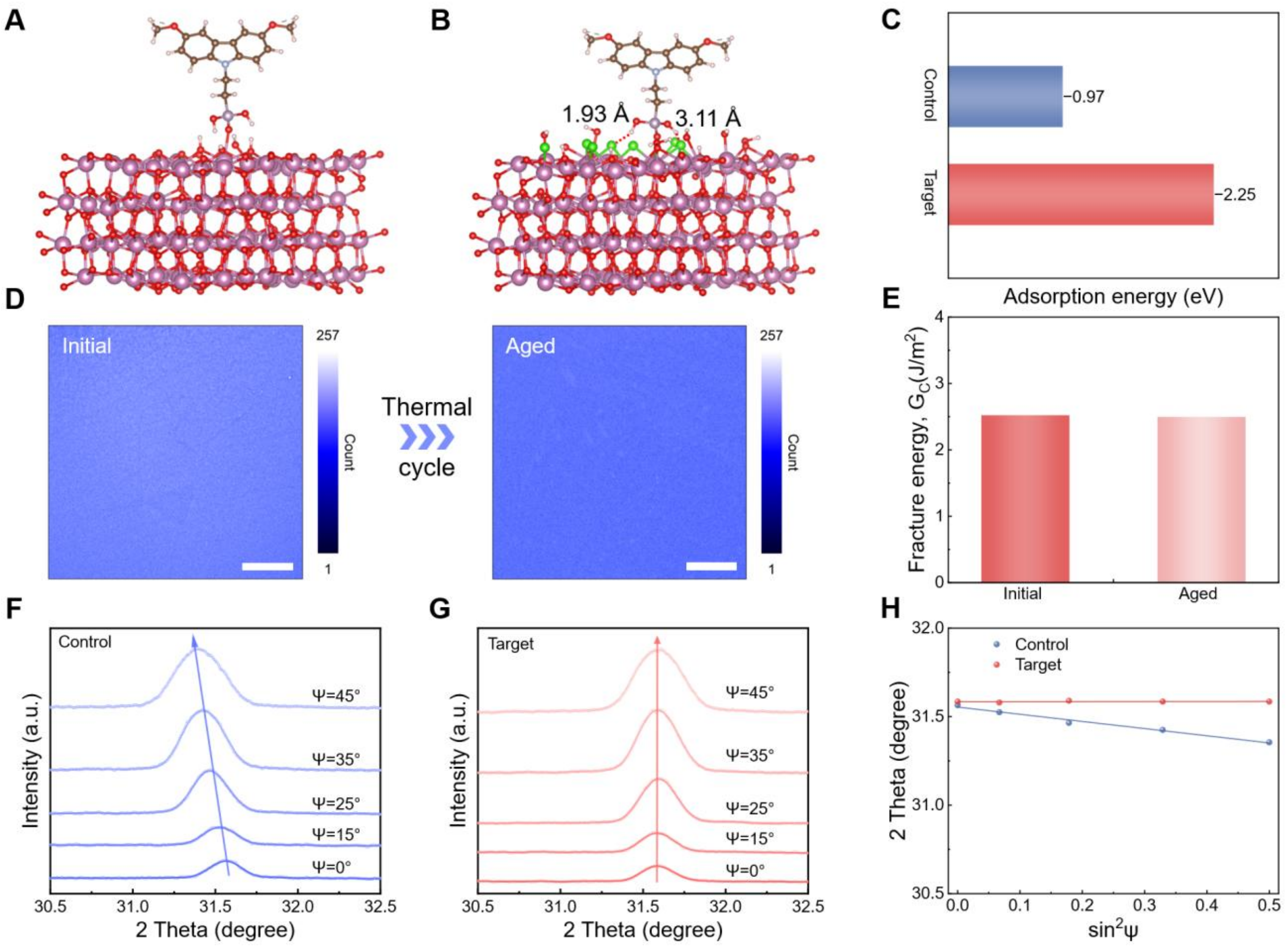

**Fig. 3. Enhancing SAM anchoring stability under thermal cycling.** Optimized adsorption configurations of the MeO-2PACz molecule on the (**A**) control and (**B**) target substrates. (**C**) Calculated adsorption energies of MeO-2PACz on the pristine and target substrates. (**D**) PL mapping images of the SAM films on the target substrate before and after thermal cycling. (**E**) Interfacial fracture energy extracted from DCB tests for the perovskite/SAM/ITO stack of the target film in the initial and thermally aged states. GIXRD patterns of the (**F**) control and (**G**) target perovskite films before and after thermal cycling. (**H**) 2θ versus $\sin^2\psi$ plots for the perovskite films on the control and target substrates.

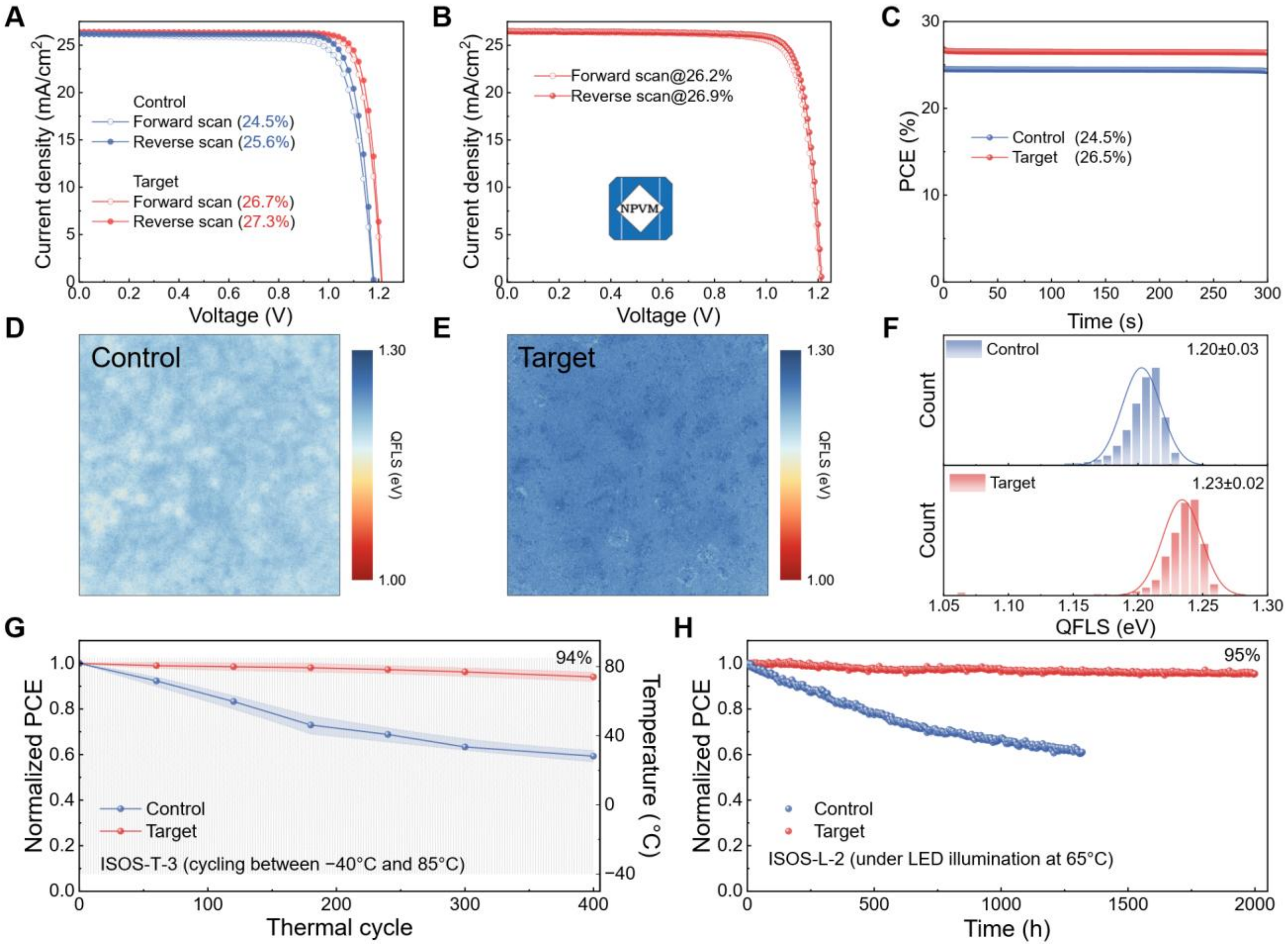

**Fig. 4. Photovoltaic performance and operational stability.** **(A)** *J-V* curves of the control and target devices measured under standard AM 1.5G illumination. **(B)** NPVM-certified *J-V* curve of the target device. **(C)** Steady-state power output of the control and target devices. QFLS mapping of the (**D**) control and (**E**) target films. **(F)** Statistical distribution of QFLS extracted from the spatial mapping. **(G)** ISOS-T-3 thermal-stress stability of the devices. **(H)** ISOS-L-2 light-soaking stability test of the devices.